\documentclass[aps, prd, onecolumn]{revtex4-2} 
\usepackage{graphicx, xcolor, hyperref}

\newcommand{\cC}{\ensuremath{\mathcal{C}}}

\newcommand{\cPT}{\ensuremath{\mathcal{PT}}}
\newcommand{\cP}{\ensuremath{\mathcal{P}}}
\newcommand{\cT}{\ensuremath{\mathcal{T}}}
\newcommand{\cK}{\ensuremath{\mathcal{K}}}
\makeatletter

\begin{document}

\title{Universality of Pattern Formation}
\author{Moses A. Schindler$^1$}\email{schindler@wustl.edu}
\author{Stella T. Schindler$^{2}$}\email{stellas@mit.edu}
\author{Leandro Medina $^1$}\email{leandro.medina.lv@gmail.com}
\author{Michael C. Ogilvie$^1$}\email{mco@wustl.edu}
\address{$^1$Physics Department, Washington University, St. Louis, MO 63130 \\ $^2$ Center for Theoretical Physics, MIT, Cambridge, MA 02139}

\begin{abstract}
 
We study a $\cPT$-symmetric scalar Euclidean field theory with a complex action, using both theoretical analysis
and lattice simulations. This model has a rich phase structure that exhibits pattern formation in the critical region.
Analytical results and simulations associate pattern formation with tachyonic instabilities in the homogeneous phase.
Monte Carlo simulation shows that pattern morphologies vary smoothly, without distinct microphases.
We suggest that pattern formation in this model may be regarded as a form of arrested
spinodal decomposition.
We extend our theoretical analysis to multicomponent $\cPT$-symmetric Euclidean scalar field theories
and show that they give rise to new universality classes of local field theories that exhibit patterned behavior in the critical region. QCD at finite temperature and density is a member of the $Z(2)$ universality class when the Polyakov loop is
used to distinguish confined and deconfined phases. This suggests the possibility of the formation of patterns
of confined and deconfined matter in QCD in the critical region in the $\mu-T$ plane.
\end{abstract}
\maketitle

\section{Introduction.} 
Reaching a theoretical understanding of finite-density QCD has been hampered for decades by a {\it sign problem}
\cite{deForcrand:2010ys,Gupta:2011ma,Aarts:2015tyj}.
The QCD sign problem originates from an imaginary term in its action induced by the chemical potential $\mu$,
giving rise to nonpositive weights in the path integral.
Sign problems are barriers to simulation in a variety of physics problems that are in some cases known to be NP hard \cite{PhysRevLett.94.170201,Barahona_1982}.
In QCD, $\mu$ explicitly breaks charge conjugation $\mathcal{C}$, but leaves the theory invariant under the combined action of $\mathcal{C}$ and complex conjugation $\mathcal{K}$. This is a form of $\cPT$ symmetry \cite{Bender:1998ke,Bender:2007nj,Meisinger:2012va}. 
$\cPT$-symmetric systems have attracted considerable theoretical and experimental interest in recent years \cite{feng, Christodoulides, Miri}; while there has been great progress in one-dimensional systems, the behaviors of $\cPT$-symmetric systems in higher dimensions are less well understood \cite{BenderBook}. We show that multifield 
$\cPT$-invariant models in higher dimensions form a rich class of systems exhibiting pattern formation around critical points. The number of experimental realizations of $\cPT$-symmetric systems is growing, and it is likely that pattern formation can be observed experimentally in engineered systems with $d\ge 2$ . Finite-density QCD is an exotic example from the Z(2) class of such models, similar to the model studied in this paper. 
This raises the possibility that finite density QCD might exhibit stable patterns consisting of the confined and deconfined phases near the critical end point. 

A quantum field theory with $\cPT$ symmetry is invariant under the combined action of a discrete linear transformation $\cP$ and an antilinear transformation $\cT$. 
For example, the $i\phi^3$ model of the Lee-Yang transition is invariant under $\cPT$ for $\cP: \phi \to -\phi$ and $\cT: i \to -i$ \cite{Fisher:1978pf}. 
In a $\cPT$-symmetric lattice theory, every transfer matrix eigenvalue must be either real or part of a complex-conjugate pair. 
While complex field theories in general suffer from a sign problem, a method was recently developed to circumvent the issue in certain $\cPT$ complex scalar theories. The procedure uses Fourier transforms to recast the action into a real dual form, which is easily simulated. The first simulations of two-component interacting scalar field models revealed patterned configurations \cite{Ogilvie:2018fov}.

Patterning behavior is observed on many length scales in physics, from condensed matter to biophysics to astronomy. 
Competition between attractive and repulsive forces of comparable magnitude is often responsible this patterning
\cite{Seul476,PhysRevLett.100.246402,PhysRevE.66.066108}. For example, it is thought that opposing nuclear and Coulomb forces may give rise to {\it nuclear pasta}  in the inner crust of neutron stars  \cite{Ravenhall:1983uh,10.1143/PTP.71.320,Caplan:2016uvu, Horowitz:2004yf}.
Notably, patterns can also form when purely repulsive forces are present \cite{Glaser_2007}.
Because imaginary coupling constants can make scalar exchange repulsive rather than attractive, there is a natural connection between $\cPT$-symmetric scalar field theory models and pattern formation. 
A patterned region of parameter space is often conceptualized in terms of {\it microphases}: subregions each characterized by a distinct type of morphology like dots, striping, or tubes. %These shapes are nontrivial, high symmetry solutions of the equations of motion. 
In the naive microphase picture, these subregions are separated from one another by a first-order transition
\cite{PhysRevE.66.066108,Ravenhall:1983uh,PhysRevE.66.066108}.

% Modified 3-17-20
%We begin by simulating a $\cPT$-symmetric scalar field theory and demonstrate that its patterned configurations exhibit smooth variation without distinct microphases as parameters are varied. 

% intro modified 8-29-2020

We study analytically a particular $\cPT$-symmetric Euclidean scalar field theory with a complex action and demonstrate that there is a region of parameter space near a phase transition where no homogeneous phase exists, due to the appearance of instabilities at nonzero momentum. 
After transformation into a form with a real, local action, simulation of the model  shows that the stable phase in this region has persistent patterning behavior, associated with those nonzero momentum instabilities. 
We give a precise criterion for the occurrence of pattern formation which generalizes to a broad class of
 $\cPT$-symmetric models.
Although different pattern
morphologies are seen as the parameters of the model are varied, the variation is smooth with no indication of
thermodynamically distinct microphases.
We connect this behavior with spinodal decomposition and nucleation.
Finally, we demonstrate that patterning is a universal phenomenon in multicomponent $\cPT$-symmetric scalar field theories. QCD at finite density is in the $Z(2)$ class of such theories, like the scalar model below. 

\section{Analytics} 
%\noindent {\bf Analytics.} 
We begin by studying the Euclidean action
\begin{equation}\label{eq:action}
S(\phi,\chi) = \sum_x \frac{1}{2}(\nabla_\mu \phi)^2 + \frac{1}{2}(\nabla_\mu \chi)^2+ V(\phi,\chi),
\end{equation}
where we set 
\begin{equation}
V(\phi,\chi) =  \frac{1}{2}m_\chi^2 \chi^2   -ig\phi\chi+ U\left(\phi\right)+h\phi.
\end{equation} 
where $U\left(\phi\right)=\lambda(\phi^{2}-v^{2})^{2}$.
Equation (\ref{eq:action}) represents a Hermitian scalar field $\phi(x)$ coupled to a $\cPT$-symmetric scalar field $\chi(x)$ by the imaginary strength $ig$. 

Our model is amenable to analytical treatment.
Because $\chi$ enters quadratically in the action $S$, it can easily be integrated out, yielding
a nonlocal effective action of the form
\begin{equation}
S_{\text{eff}}=\sum_{x} \left[\frac{1}{2}(\partial_{\mu}\phi(x))^{2}+\lambda(\phi^{2}-v^{2})^{2}+h\phi\right]\nonumber
+\frac{g^{2}}{2}\sum_{x,y}\phi(x)\Delta(x-y)\phi(y).
\end{equation}
This model has been extensively studied in the case $m_\chi = 0$;
see, {\it e.g.} \cite{PhysRevE.66.066108} and references therein.
In the two-dimensional case, pattern formation with stripes and dots is
known to occur.
The $m_\chi = 0$ limit is sometimes described in the condensed matter literature as Coulomb frustrated because the extra interaction acts against the symmetry-breaking behavior of the  $\phi^4$ model 
\cite{PhysRevE.66.066108,PhysRevLett.100.246402,ORTIX2009499}.
However, our simulations show that the observed patterning behavior is not tied to the long-range nature of the Coulomb interaction, and also occurs for a Yukawa interaction, {\it i.e.}, when $m_{\chi} \ne 0$.

% modified 5-20-20
We determine the value of the order parameter $\phi_0$ at tree level by minimizing the potential or equivalently by minimizing the effective potential associated with $S_{\text {eff}}$:
\begin{equation}
V_{\text{eff}}\left(\phi_0\right)=\lambda\left(\phi_0^2-v^2\right)^2+g^2\phi_0^2/2 m_\chi^2-h\phi_0. \label{eq:v-eff}
\end{equation}
The effect of $\chi$ on $\phi_0$ for $h=0$ is to restore the symmetric value $\phi_0 = 0$  at sufficiently large
values of $g$. The value of $\phi_0$ is essentially the expected value of the zero-momentum component of the field 
$\phi$. Pattern formation is associated with Fourier modes $\tilde\phi\left(q\right)$ with nonzero $q$.
% added 8-3-2020
One approach to understanding pattern formation is to expand $S_{\text {eff}}$ in a derivative expansion. The last, nonlocal term in $S_{eff}$ generates the terms
\begin{equation}
\int d^4 x {g^2 \over 2 m^2}\phi\left(x\right)\sum_{n=0}^{\infty}  \left( {-\nabla^2 \over m^2 }\right)^n 
\phi\left(x\right).  
\end{equation}
The $n=0$ term in this expansion is the last term in $V_{\text {eff}}$.
Crucially, the $n=1$ term is negative, indicating that the quadratic derivative term becomes negative for sufficiently large $g$. When the quadratic kinetic term is sufficiently negative, the homogeneous phase is unstable to perturbations with nonzero wave number.  Thus the occurrence of a pattern-forming region is a manifestation of a Lifshitz instability \cite{cha95}.

Additional information on the phase structure follows from the inverse $\phi$ propagator obtained from $S_{\text {eff}}$ at tree level:
\begin{equation}
G^{-1}\left(q^2\right)=q^2 +U''(\phi_0)+{{g^2} \over{q^2+m_\chi^2}}
\end{equation}
where $U''(\phi_0)=-4\lambda v^2 +12\lambda \phi_0^2$.
The allowed phases of the model are determined using $\phi_0$ and $G\left(q\right)$.
The poles of the propagator are obtained from the zeros of
$(q^2+m^2_\chi)(q^2+U)+g^2$.
This quadratic in $q^2$ has real coefficients, so its roots $r_1$ and $r_2$ are either both real or form a complex conjugate pair.
If both zeros occur at $q^2<0$, then the propagator must decay exponentially. We refer to this region of parameter space as the normal region. If the zeros are complex, they must form a complex conjugate pair with $r_1=r_2^*$, and the propagator decays exponentially with sinusoidal modulation. We refer to this region of parameter space as the complex region.   In both of these cases, the homogeneous phase is stable against small fluctuations for all values of $q^2>0$. The boundary between these two behaviors is by definition a disorder line \cite{PhysRevB.1.4405}. If one of the roots is positive and the other negative, then the inverse propagator $G^{-1}\left(q\right)$ is negative at $q^2=0$, and $\phi_0$ is not a stable solution. This is the unstable region of parameter space, which occurs in the case $g=0$ when $V''(\phi_0)<0$.
If both roots are positive, then $\phi_0$ is stable against perturbations at $q^2=0$, but unstable in a linearized analysis against perturbations with $q^2=r_1$ or $q^2=r_2$. This is the region we identify as the pattern forming region.

The unstable region, where $\phi_0$ is unstable against $q^2=0$ perturbations, is determined simply by
$V''_{eff}(\phi_0)=U''(\phi_0)+g^2/m^2_\chi<0.$
Note that as $g^2\rightarrow 0$, we obtain the standard result $V''(\phi_0)<0$ for the unphysical region.
A more comprehensive analysis requires the roots
$r_1$ and $r_2$ of $G^{-1}\left(q^2\right)$. 
We see immediately that if one root is positive and the other negative,
the homogeneous solution $\phi_0$ is unstable.
This is precisely the condition $V''_{eff}<0.$
The roots are determined from the quadratic formula and given by
\begin{equation}
r_{1,2}={1\over 2}\left[-(m^2_\chi+U'')\pm\sqrt{(m^2_\chi+U'')^2-4 m^2_\chi U'' -4g^2}\,\right]
\end{equation}
which can be rewritten immediately as
\begin{equation}
r_{1,2}={1\over 2}\left[-(m^2_\chi+U'')\pm\sqrt{(m^2_\chi-U'')^2 -4g^2}\,\right]
\end{equation}
The complex pole region is determined by
$(m^2_\chi-U'')^2 -4g^2<0$ so the boundary of the complex region is given by $(m^2_\chi-U'')^2 -4g^2=0$.
The stability condition $U''(\phi_0)+g^2/m^2_\chi>0$ is automatically satisfied in this region.
The parameters $g$ and $m_\chi^2$ are always positive, but $U''$ is negative for small $\phi$ and positive for large 
$\phi$. Thus there are two possible solutions:
$2g=m^2_\chi-U''$
when $U''<m^2_\chi$, and 
$2g=U''-m^2_\chi$
when $U''>m^2_\chi$.
The normal region has $r_{1,2}<0$ while the pattern-forming region has $r_{1,2}>0$.
In either case, the discriminant of the quadratic formula must be positive.
From the solution for the roots we see that the stability condition $U''(\phi_0)+g^2/m^2_\chi>0$ 
implies that for real roots, the discriminant in the quadratic formula is always smaller in magnitude than
 $\left| m_\chi^2+U''(\phi_0)\right|$. The normal region is thus  characterized by $m_\chi^2+U''(\phi_0)<0$
 in addition to the discriminant  condition $(m^2_\chi+U'')^2-4 m^2_\chi U'' -4g^2>0$ and the stability condition
 $U''(\phi_0)+g^2/m^2_\chi>0$. The pattern forming region is characterized by $m_\chi^2+U''(\phi_0)>0$
 in addition to the discriminant and stability conditions.
 
\begin{figure}
\includegraphics[width=4.5in]{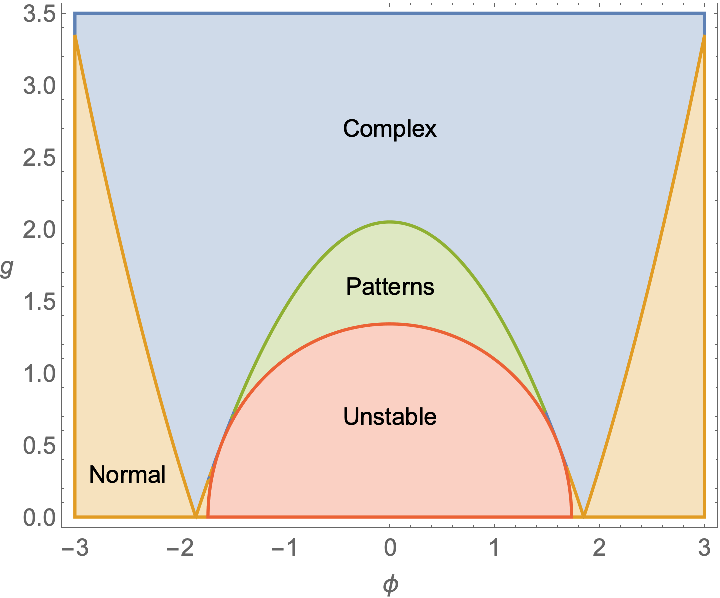}
\caption{Phase diagram of model in the $n$-$g$ plane for the parameter set $m^2=1/2$, $\lambda=1/10$ and $v=3$. In the region labeled Complex (blue), the poles as a function of  $q^2$ in the $\phi$ propagator are both complex. In the Normal region (orange), both poles are real and negative. The region labeled Patterns (green) is where both poles are real and positive. It is in this region where persistent patterns occur. In the Unstable region (red), both poles are real with one positive and one negative. This region is not thermodynamically stable. }
\label{fig:phase-diagram}
\end{figure}

In Figure \ref{fig:phase-diagram}, we plot the regions for the four distinct phases of the model as a function of
$\left< \phi \right>$ for the parameter set $m^2=1/2$, $\lambda=1/10$ and $v=3$. The different regions are classified by the nature of the poles of the $\phi$ propagator in the $q^2$ complex plane. We denote the region of parameter space where both poles are real and negative as ``Normal'' (orange in \ref{fig:phase-diagram}). This leads to exponential decay of the $\phi$ propagator, as it does in conventional field theories. In the region labeled Complex (blue), the poles as a function of  $q^2$ in the $\phi$ propagator are complex conjugates. This region is similar to the so-called broken $\cPT$ region of $\cPT$-symmetric quantum mechanical models. The $\phi$ propagator in this region also decays exponentially, but with sinusoidal modulation. The boundary between the Normal and Complex regions is called a disorder line. The region labeled Patterns (green) is the region where both poles are real and positive; it is in this region where persistent patterns occur.  In the Unstable region (red), both poles are real with one positive and one negative. This region is not thermodynamically stable, and is inaccessible as an equilibrium state in simulations in the canonical ensemble. As in the familiar case of a ferromagnetic Ising model, the phase diagram has a cut at $h=0$ across which $\phi_0$ jumps. The two sides of the cut may be in the Normal, Complex or Pattern regions. We see evidence for this behavior in the Pattern region in our lattice simulations at $h=0$: $\phi_0$ can have either sign, and the observed patterns are correspondingly inverted.

The effective action $S_{\text{eff}}$ is a function of $g^2$, and can be continued to $g^2<0$. This corresponds to the continuation $g \rightarrow ig$ in $S$, in which case the action $S$ becomes real, and neither pattern formation nor complex $q^2$ poles can occur. The small areas of normal behavior seen in Fig. \ref{fig:phase-diagram} in between the complex and unstable phases are connected to the larger normal region when the phase diagram is plotted as a function of $g^2$ and $g^2<0$ values are included. The relative size of the unstable and pattern-forming regions is directly controlled by $m_\chi$. The Coulomb-frustrated model is obtained in the limit $m_\chi \rightarrow 0$. In that limit, the boundary between the unstable and pattern-forming region moves to $g=0$, and the unstable region is seen only for $g^2<0$. In the limit $m_\chi \rightarrow \infty$, the pattern-forming and complex regions vanish and standard $\phi^4$ behavior is obtained.

%The zeros of $G^{-1}$ determine the stability of the homogeneous phase.
%If both zeros occur at $q^2<0$, then the propagator must decay exponentially. If the zeros are complex, they must form a complex conjugate pair, and thus the propagator decays exponentially with sinusoidal modulation. In both these cases, the homogeneous phase is stable. The boundary between these two behaviors is by definition a disorder line \cite{PhysRevB.1.4405}. 
%If one or both zeros are real and positive, then the linearized theory predicts that such modes will grow exponentially.
%indicating that the homogeneous phase is unstable. This is where pattern formation occurs.
% modified 3-17-20 to explain what we mean by tachyonic
%We predict the pattern-forming region analytically
%by determining the region where the homogeneous phase is stable
%with respect to zero-momentum fluctuations but unstable with respect to
%fluctuations of nonzero momentum.
 
The inverse propagator $G^{-1}(q)$ has a minimum at  $q^2>0$ provided $g > m_\chi^2$.
The propagator has poles with $q^2>0$ if the minimum of $G^{-1}(q)$ lies below zero; that is, when $2g-m_\chi^2 -4\lambda v^2 +12\lambda \phi_0^2<0$. Such poles are tachyonic in the usual sense of continuation to Minkowski space and are
generally taken to indicate an instability of the system such that
a homogeneous equilibrium phase is not stable.
The two-point function will not decay with spatial separation, instead exhibiting oscillatory behavior.
%This is closely related to the appearance of spatially modulated phases 
%in systems with higher-derivative terms, as first described by Lifshitz \cite{cha95}.
This tachyonic region is in reasonable agreement with the boundaries of the pattern-forming region observed in simulation, subject to the limitations imposed by lattice size and spacing.

An equivalent approach to determining the phase structure is to start from $S$ rather than $S_{\text{eff}}$ and find the static solution
$\left(\phi_0,\chi_0\right)$  which minimizes $V(\phi,\chi)$. Unless the underlying $\cPT$ symmetry of $S$ is broken,
$\phi_0$ will be real and $\chi_0$ will be purely imaginary. Linearizing the propagator around the static solution, we find the inverse propagator for the $\left(\phi,\chi \right)$ set of fields is $q^2+\mathcal{M}$, where $\mathcal{M}$ is the $2\times 2$ mass matrix
\begin{equation}
\mathcal{M}=\left(\begin{array}{cc}
\frac{\partial^{2}V}{\partial\phi^{2}} & \frac{\partial^{2}V}{\partial\phi\partial\chi}\\
\frac{\partial^{2}V}{\partial\phi\partial\chi} & \frac{\partial^{2}V}{\partial\chi^{2}}
\end{array}\right),
\end{equation}
which is
\begin{equation}
{\mathcal M}=\left(\begin{array}{cc}
U''\left(\phi_0\right) & ig\\
ig & m_{\chi}^{2}\\
\end{array}\right).
\end{equation}
The mass matrix $\mathcal M$ is not Hermitian, but it satisfies a $\cPT$ symmetry condition 
\begin{equation}
M = \sigma_3 M^* \sigma_3.
\end{equation}
This condition implies that the eigenvalues of $\mathcal M^*$ must be the same as those of $\mathcal M$, and thus they are either both real or form a complex pair. The zeros of the inverse matrix propagator can be obtained as the zeros of the characteristic equation
\begin{equation}
\det\left(q^2 + {\mathcal M}\right) = \left(q^2\right)^2+\textrm{ tr}\left({\mathcal M}\right) q^2 + \det\left({\mathcal M}\right).
\end{equation}
The coefficients $\textrm{ tr}{\mathcal M}$ and $\det\left({\mathcal M}\right)$ are real, implying that the roots are either both real or form a complex conjugate pair. The zeros of the characteristic equation are the propagator poles and so the two methods give the same results.

%(*arrays contain averages of all fields, not just field #0 as we \
%want, hence we take each '1 mod 3'th element*)
%(*configuration \
%information: 32x32 lattice, 400 sweeps for thermalization, 1500 \
%configuration snapshots with 250 sweeps in between each *)
%(*g values \
%were 0.76, 0.89*)
%(* (h/c)(a/b) stands for hot/cold and above/below \
%(the critical point), e.g. "ha2" is a hot run with g=0.89*)

% Modified 3-17-20
%\noindent {\bf Simulations.} 
\section{Simulations} 
The complex Euclidean action $S$ may be cast into a real local form suitable for simulation
by the following steps. First, we write the derivative term for $\chi$ as a functional integral over a momentum 
$\pi_\mu$ conjugate to $\nabla_\mu \chi$:
%\begin{equation}\label{eq:trick1}
%\exp\left[-\sum_x \frac{1}{2}(\nabla_\mu \phi)^2\right]=
%\int \left[d\chi\right] \exp\left[-\sum_x \frac{1}{2}(\pi_\mu \phi)^2+i\pi_\mu \nabla_\mu\chi \right].
%\end{equation}
\begin{equation}\label{eq:trick1}
\exp\left[-\sum_x \frac{1}{2}(\nabla_\mu \chi)^2\right]=
\int \left[d\pi_\mu\right] \exp\left[-\sum_x {1 \over 2} \pi_\mu^2+i\pi_\mu \nabla_\mu\chi \right].
\end{equation}
After integration by parts on the $i\pi_\mu \nabla_\mu\chi$ term,
the functional integral over $\chi$ becomes Gaussian and local.
After integration over $\chi$,
we obtain a dual action of the form \cite{Ogilvie:2018fov}
\begin{equation}\label{eq:dual}
\tilde{S} = \sum_x \frac{1}{2}[\nabla_\mu\phi(x)]^2 + \frac{1}{2}\pi_\mu^2(x) + \tilde{V}[\phi(x),\nabla\cdot\pi(x)],
\end{equation}
where the dual potential $\tilde{V}$ is given by $\tilde{V}\left(\phi,\nabla\cdot\pi \right)={\left(\nabla\cdot\pi-g\phi\right)^2}
/{2m^2_\chi}+\lambda(\phi^{2}-v^{2})^{2}+h\phi$.
The dual action $\tilde{S}$ is real so the sign problem has been eliminated. The dual action is easily simulated using the Metropolis algorithm to update $\phi$ and $\pi_\mu$.
The field $\phi$ is the order parameter associated with the $Z(2)$ symmetry, and we focus on its behavior.
Patterns are observed in this model at $h=0$ in two and three dimensions \cite{Ogilvie:2018fov}. 
We have performed extensive Monte Carlo simulations on $32^{2}$ and $64^{2}$ lattices varying the parameters $g$ and $h$ holding fixed the other parameters at $m_{\chi}^{2}=0.5$, $\lambda=0.1$ and $v=3$. These are the same parameters used in in Fig. \ref{fig:phase-diagram}.
We observe similar behavior in the $\left( g, h \right)$ plane for $d=3$. 

%Here we analyze Eq. (\ref{eq:action}) over the full $g-h$ plane, with simulations performed on a $64^{2}$ lattice with parameters $m_{\chi}^{2}=0.5$, $\lambda=0.1$ and $v=3$. We have observed similar behavior in the $g-h$ plane for $d=3$.

\begin{figure}
\includegraphics[width=4.5in]{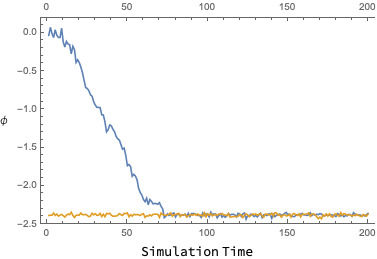}
\caption{Spatial average of $\phi$ versus Monte Carlo time for $g=0.76$ using the parameter set $h=0$,
$m^2=1/2$, $\lambda=1/10$ and $v=3$ on a $32\times32$ lattice. Measurements were begun after $400$ lattice sweeps, and there are $250$ sweeps in between each measurement. The first $200$ measurements are shown.  The hot start is in blue, and the cold start is orange.}
\label{fig:MCtime_g0.76}
\end{figure}

\begin{figure}
\includegraphics[width=4.5in]{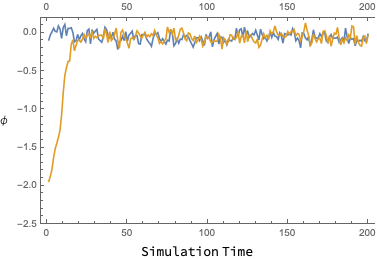}
\caption{Spatial average of $\phi$ versus Monte Carlo time for $g=0.76$ using the parameter set $h=0$,
$m^2=1/2$, $\lambda=1/10$ and $v=3$ on a $32\times32$ lattice. Measurements were begun after $400$ lattice sweeps, and there are $250$ sweeps in between each measurement. The first $200$ measurements are shown. The hot start is in blue, and the cold start is orange.}
\label{fig:MCtime_g0.83}
\end{figure}

We have checked for the onset of the pattern forming region as $g$ is increased at $h=0$. Figures \ref{fig:MCtime_g0.76} and \ref{fig:MCtime_g0.83} show the evolution of the spatial average of $\phi$ under the Monte Carlo algorithm as a function of the number of measurements. Measurements were begun after $400$ lattice sweeps, where each sweep represents a Monte Carlo update of every site on the lattice. There are $250$ sweeps in between each measurement; the first $200$ measurements are shown. Results are shown in both cases for cold (ordered) and hot (random) initial conditions. We see very clearly that at $g=0.76$  both the cold start and the hot start are converging to an equilibrium behavior consistent with $\left< \phi \right>\ne 0$. On the other hand, at $g=0.83$  both the cold start and the hot start are converging to an equilibrium behavior consistent with $\left< \phi \right> =0$. Configuration snapshots indicate that pattern formation begins at a critical value $g_c$ in the range $0.76 \le g_c \le 0.83$. Although the convergence of the hot start at $g=0.76$  to the ordered phase is perhaps three times slower than the convergence of the cold start  to the patterned phase at $g=0.83$, the times required are much shorter than the length of our runs.

\begin{figure}
\includegraphics[width = 3.5 in]{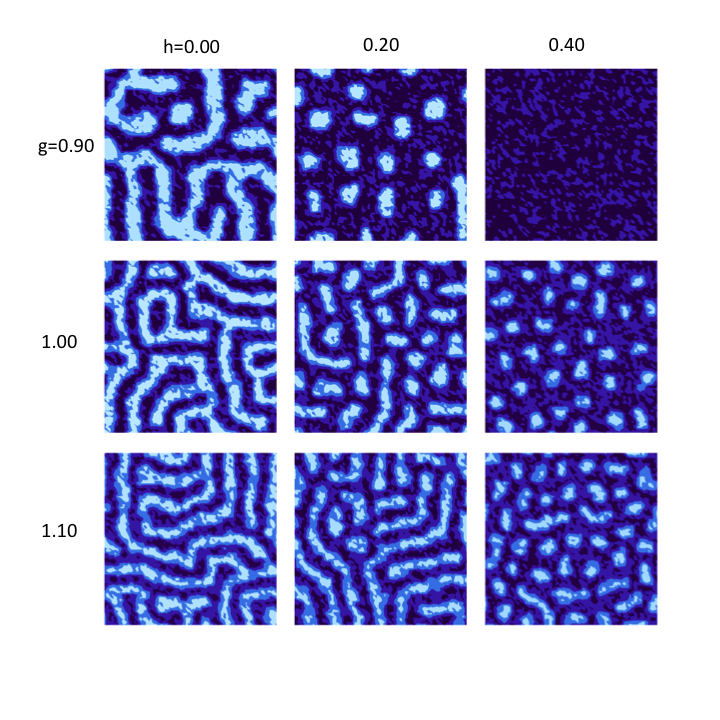}
\caption{Configuration snapshots of $\phi$ in Eq. (\ref{eq:dual}) on a $64^{2}$ lattice for several values of $h$ and $g$. 
The color scale runs from $-3$ to $3$, ranging from dark to light.
From left to right, $h=0.0,\,0.2,$ and $0.4$; from top to bottom: $g=0.9,\,1.0,$ and $1.1$.
The other parameters are $m_{\chi}^{2}=0.5$, $\lambda=0.1$, and $v=3$.
Each configuration was obtained after a hot start followed $20,000$ sweeps through the lattice.}\label{fig:config}
\end{figure}
 
Figure \ref{fig:config} shows nine configuration snapshots of $\phi$ from a $64^2$ lattice, each taken after $20,000$ lattice sweeps, in which the fields at every site are updated. 
These snapshots are taken from a large dataset that
extends from $g=0.0$ to $g=2.0$  at $h=0$ out to $g$ around $1.0$ at  $h=0.95$, covering the region where pattern formation occurs. For smaller values of $g$, the length scale of pattern formation is too large for a $64^2$ lattice to reveal much information. For larger values of $g$, the length scale of pattern formation is on the order of the lattice spacing so the details of any pattern formation are lost. 

Though each configuration shows a distribution of different shapes, we observe a number of general features. 
At $h=0$ and intermediate values of $g$, we primarily see long curved line segments, often called {\it stripes}.
For a given pair of $g$ and $h$ values, the width of these stripes is fairly uniform,
but there is considerable randomness in the overall pattern. 
As $h$ increases for fixed $g$, the typical line segment length decreases until commensurate with the width, forming a {\it dot}. 
The dots then shrink until disappearing at the boundary of the pattern-forming region. 
As $g$ increases for fixed $h$, the width of the morphologies tends to decrease. 
The variation in pattern morphologies is smooth as $g$ and $h$ are varied, 
as are the change in histograms of $\phi$ values and the average action $\langle S \rangle$. 

\begin{figure}
\includegraphics[width = 3.5 in]{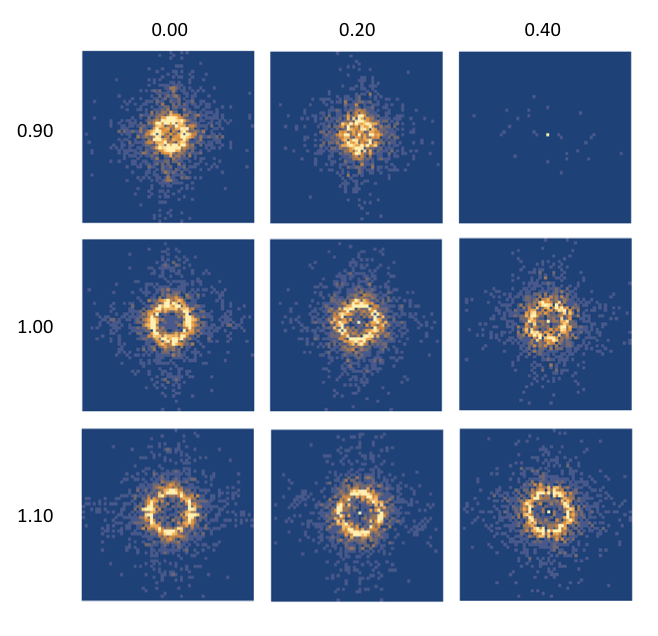}
\caption{The absolute value of the Fourier transforms $\tilde{\phi}\left(k\right)$ for the configurations shown in Fig.\ \ref{fig:config}. We scale each graph do that its colors run from 0 (dark blue) to 10 (light yellow). Any lattice point with magnitude greater than 10 is set to 10. The ring-shaped Fourier transforms correspond to patterning in Fig.\ \ref{fig:config}.}\label{fig:fourier}
\end{figure}

Figure \ref{fig:fourier} shows the absolute value $| \tilde\phi |$ of the Fourier transforms of the configurations $\phi$ shown in figure \ref{fig:config}. All graphs use the same color scale, but we clip large values for visual clarity. This is necessary because as $h$ increases, the zero mode $|\tilde\phi(0)|$, representing the expected value of $\phi$, dominates. The eight patterned configurations have a common feature: a ring in momentum space. The radius of the ring increases with $g$ but is not heavily dependent on $h$. 
At $\left( g,h \right)=\left(0.9, 0.4\right)$, there is no ring and 
$| \tilde\phi ( 0 ) |$ is large. This corresponds to a nonzero expectation value and the absence of patterns.
The tachyonic region observed in simulations is in reasonable agreement with the pattern-forming region obtained
analytically, given the limitations imposed by lattice size and spacing.

In most of the two-dimensional simulations, we see a fairly complete ring in momentum space, consistent with pattern formation without preferred directions. In some cases, however, a smaller number of modes on the ring are excited, and the absence of isotropy is evident in the configurations. This is most common near the boundary of the pattern-forming region. This may be related to finite size effects, to lattice pinning or to some form of locking into an atypical region of configuration space. It is known that the ground state for Ising systems with a long-range frustrating interaction is the striped phase \cite{CMP347.983}, 
but we are not aware of a similar result for the frustrated Yukawa system with continuous spins.

% ORIGINAL- MODIFIED 2-25-20
%It is of course still possible that some currently unknown operator might serve as an order parameter for what are referred to as {\it geometric transitions} associated with percolative behavior. It is known in the case of the d=1 Ising model that there is an infinite class of nonlocal string operators, each with its own disorder line. This is associated with the behavior of the model in an imaginary magnetic field, a
%$\cPT$-symmetric system \cite{PhysRevE.96.062123}, 
%so it is plausible that such behavior may exist in other $\cPT$-symmetric theories.
%Such transitions need not affect thermodynamic behavior.

\section{Discussion}
\begin{figure}
\includegraphics[width=1.1in]{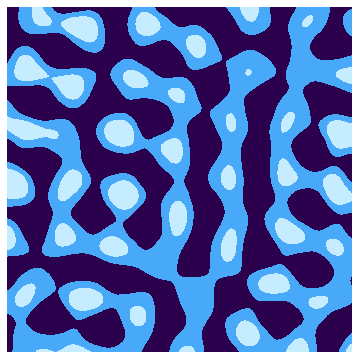} \includegraphics[width=1.1in]{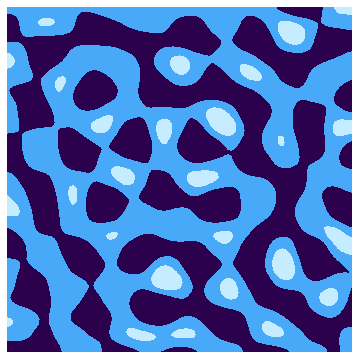}
\caption{(color online) Topography of synthetic configurations of the form Eq. (5), for two different sets of $(k_j, \delta_j)$ and $\phi_0 = 0$. The light blue, turquoise, and dark blue regions correspond to the regions where $\phi(x) > 4$, where $0 < \phi(x) < 4$, and where $\phi(x) < 0$, respectively. (This is equivalent to plotting the regions of $\phi(x)$ that are positive when $\phi_0 = -4$ or $0$.) The synthetic model topography mimics the patterns in the figure 1 field configurations: the light blue region forms a set of droplets, and the turquoise region looks like a striped configuration.}
\label{fig:synthgraphs}
\end{figure}

As a first step towards approximating the behavior of the equilibrium patterning state, 
let us consider a simple model that provides insight into the change between different patterning behaviors. 
Motivated by the Fourier-space simulation results, we consider configurations of the form
\begin{equation}\label{eq:synthetic}
\phi\left( x \right) = \phi_0 +\sum_{j=1}^n A \cos \left( k_j \cdot x  -\delta_j \right),
\end{equation}
where the momenta $k_j$ are constant in magnitude but uniformly distributed in direction;
the phases $\delta_j$ are also uniformly distributed.
Fig. \ref{fig:synthgraphs} shows the topography of two configurations with different sets of $k_j$ and $\delta_j$ values. This rather simple model reproduces much of the observed pattern morphology. It is clear that any configuration with the structure of Eq. (\ref{eq:synthetic}) will tend to produce topographic contours that appropriately mimic the patterns of stripes and droplets as $\phi_0$ is varied. The observed pattern morphology varies continuously with $\phi_0$ suggesting  that there is no phase transition between different microphases when this picture is valid.
Conceptually, we expect that the amplitude $A_j$ for each component of the sum is determined by the nonlinear dynamics, but the phases $\delta_j$ represent collective coordinates associated with would-be zero modes of the system.
However, we note that reproducing all aspects of equilibrium patterned behavior is a difficult problem. 

%The equilibrium patterning behavior observed in our simulations is strikingly similar to that in phase transition dynamics.
%We can model the dynamics of $\phi$ by a Langevin equation
%\begin{equation}
%{\partial \phi( x)\over \partial t}=-\Gamma {\delta S_{\text{eff}} \over \delta \phi(x)}+\eta(x).
%\end{equation} 
%where as usual $\Gamma$ is a decay constant and $\eta(x)$ is a white noise term.
%The difference between this model and a standard $\phi^4$ field theory is the nonlocal term in $S_{\text{eff}}$
%induced by $\chi$, which stabilizes $\left< \phi \right>$ in what would otherwise be an unstable region of the phase diagram.
%It is easy to show that the dynamics of pattern formation have the same enhanced modes in momentum space as the  equilibrated configurations do. 
%The dichotomy between a tachyonic origin of patterned phases and the microphase model
%is reminiscent of the distinction between spinodal decomposition and nucleation and growth
%in phase transition dynamics. Spinodal decomposition is the mechanism by which
%unstable states equilibrate while nucleation and growth is associated with the decay of
%metastable states.  It is known that there is typically no sharp boundary 
%between these two mechanisms in phase transition dynamics \cite{PhysRevA.11.1417,Binder1978,PhysRevB.45.9620}.
%Because of the close connection between dynamics and statics in this model, we propose that the relation of our tachyonic picture
%to the microphase picture is essentially the same as that of spinodal decomposition to nucleation and growth.

The numerical and analytical results taken together suggest that the patterned region is a single thermodynamic phase. 
Patterned morphologies change smoothly into one another as $g$ and $h$ vary, exhibiting a common ring-shaped form in Fourier space.
We see no evidence for the existence of distinct microphases, 
nor for first- or second-order thermodynamic transitions between supposed microphases. 
Moreover, the analytics suggest that all patterning arises from a common origin, the tachyonic modes. 
It is possible that some currently unknown operator might serve as an order parameter for what are referred to as {\it geometric transitions}. In the case of the d=1 Ising model that there is an infinite class of nonlocal string operators, each with its own disorder line. This is associated with the behavior of the model in an imaginary magnetic field, a
$\cPT$-symmetric system \cite{PhysRevE.96.062123}, so such behavior may exist in other $\cPT$-symmetric theories.
Such transitions need not affect thermodynamic behavior. 
Another possibility is that the existence and meaning of microphase concept may be closely related to concepts from
phase transition dynamics, to which we now turn.

\subsection{Model A dynamics} 
%\noindent {\bf Model A dynamics.} 
It is enlightening to consider the dissipative dynamics of our model
as compared with similar behavior seen in phase transition dynamics of closely related systems. 
\cite{RevModPhys.65.851,doi:10.1080/00018739400101505,cha95}.
Because the order parameter $\phi$ is not conserved, we can model its dynamics using $S_{\text{eff}}$ and
model A dynamics, {\it i.e.}, with a Langevin equation
\begin{equation}
{\partial \phi( x,t)\over \partial t}=-\Gamma {\delta S_{\text{eff}} \over \delta \phi(x,t)}+\eta(x,t).
\end{equation} 
where as usual $\Gamma$ is a decay constant and $\eta$ is a white noise term.
The dependence on Langevin "time" is suppressed for notational simplicity.
The difference between this model and the standard model A dynamics of a $\phi^4$ field theory obtained at $g=0$ 
is the nonlocal term in $S_{\text{eff}}$ induced by $\chi$. 
If we linearize the Langevin equation around a homogeneous solution $\phi_0$ and transform to momentum space, it becomes
\begin{equation}
{\partial \tilde\phi( q,t)\over \partial t}=-\Gamma \left(q^2+U''(\phi_0)+{g^2 \over q^2+m_\chi^2}\right)\tilde\phi(q,t)+\tilde\eta(q,t)
=-\Gamma G^{-1}(q)\tilde\phi(q,t)+\tilde\eta(q,t)
\end{equation} 
If the nonlocal term were not present, $\phi_0$ would be unstable when $U''\left(\phi\right)<0$, and spinodal decomposition would occur.
Comparing the dynamics when $g\ne 0$ to the purely local model when $g=0$, we see that the large-$q$ behavior
is identical, but $g\ne 0$ changes the small-$q$ behavior. In the region where pattern formation occurs in equilibrium, model A dynamics is that of spinodal decomposition  for large $q$, but relaxational for small $q$. 
This suggests that the equilibrium patterning behavior of this model may be understood as a form of arrested spinodal decomposition.
Starting from a homogeneous solution $\phi_0$ with $U''\left(\phi\right)<0$ , the early-time Langevin evolution will produce the exponentially growing modes
of spinodal decomposition for large $q$, but for small $q$ fluctuations are damped. Spinodal decomposition is arrested at a characteristic scale
in momentum space, with $\phi_0$ stabilized by the nonlocal term.
From this point of view, the chief dynamical difference between the patterned region and the unstable region is that in the patterned region, low $q$
fluctuations are suppressed but grow exponentially with $t$ in the unstable region.

The presence of a characteristic scale, obvious in the Fourier transforms of configuration snapshots,
may seem reminiscent of model B dynamics for a conserved order parameter \cite{RevModPhys.65.851,doi:10.1080/00018739400101505,cha95}.
In model B dynamics with $g=0$, the system will
eventually leave the spinodal region $U''\left(\phi\right)<0$ as phase separation completes.
This does not happen in this model when $g\ne 0$; large homogeneous regions are unstable to
fluctuations of nonzero $q$.
In particular, coarsening at arbitrarily large scales does not occur: the order parameter is not conserved,
and large-scale (small-$q$) fluctuations relax quickly.

\subsection{Arrested spinodal decomposition vs. microphases}

The dichotomy between a tachyonic origin of equilibrium patterned phases and the microphase model
is reminiscent of the distinction between spinodal decomposition and nucleation and growth
in phase transition dynamics. If we take $g=0$ in $S_{eff}$, we recover standard model A dynamics
for a $\phi^4$ field theory.
A homogeneous field configuration $\phi_u$ with $U''(\phi_u)<0$ will decay via spinodal decomposition,
characterized by time-dependent patterns characterized by coarsening and phase separation.
In model A dynamics, the field is unstable against exponentially growing low-momentum modes,
with the  $q=0$ growing fastest.
On the other hand, a metastable field configuration $\phi_m$ is characterized as a local minimum of $U$,
but not the global minimum $\phi_g$ of $U$, so that $U(\phi_m)>U(\phi_g)$.
It is easy to produce such metastable states in a $\phi^4$ model by adding a symmetry-breaking term
$-h\phi$ to the action, as we have done.
Conceptually, spinodal decomposition is naturally understood in momentum space, while nucleation
is understood in real space. Classically, they were considered to be two different processes.
However, it is known that there is typically no sharp boundary 
between these two mechanisms observed in simulations of phase transition dynamics \cite{PhysRevA.11.1417,Binder1978,PhysRevB.45.9620}.

In the microphase approach to pattern formation \cite{Ravenhall:1983uh,PhysRevE.66.066108},
solutions of the field equations are found which represent idealized geometries such as dots or stripes,
from which different microphases are constructed. The microphase with the lowest free energy
is considered to be the stable phase, and in principle there are first-order phase transitions between distinct
microphases. This strategy is explored extensively in \cite{PhysRevE.66.066108} for the Coulomb-frustrated
$\phi^4$ model. It should be clear that the stable microphases are closely related to the bubbles of nucleation theory,
in the same way that arrested spinodal decomposition is related to spinodal decomposition.
However, we do not see any evidence for a phase transition between microphases  in our lattice simulations.
The pattern morphology seen in simulations at different values of $h$ and $g$ does not show abrupt changes,
as the microphase picture predicts. On the other hand, there are models
for which it has been rigorously shown that the zero-temperature behavior is that
of an ordered striped phase \cite{PhysRevB.84.144402}, as suggested by the microphase picture.
We propose that the relation of our tachyonic picture of pattern formation
to the microphase picture is essentially the same as that of spinodal decomposition to nucleation and growth.
There are regions of parameter space where the behavior of a patterning system are well-described
by microphases, or by arrested spinodal decomposition, but the distinction is not sharp.

\subsection{Computational complexity}
The observation of pattern formation in field theories with complex actions raises interesting issues about 
computational complexity as well as ergodicity breaking in bosonic models with sign problems. 
Some characteristics observed in our simulations, such as large numbers of distinct configurations
and slow quasizero modes, are reminiscent of glassy behavior, and it has been argued that at low temperatures, the patterned phase gives way to a glassy phase \cite{PhysRevLett.85.836,PhysRevB.64.174203}
It is known that the problem of finding the ground state of an Ising model with general couplings is 
NP-hard \cite{Barahona_1982}.
Certain fermionic models with sign problems have been mapped to the Ising spin glass, a known NP-hard problem 
\cite{PhysRevLett.94.170201}.
We are unaware of general results for bosonic models with sign problems.
However,  the Coulomb frustrated
model, which our model becomes in the $m_\chi \rightarrow 0$ limit, has been used
as an example of a system exhibiting glassy behavior without quenched disorder \cite{PhysRevLett.85.836,
PhysRevB.64.174203}.

\section{Universality}
We can extend conventional scalar field theory universality classes by augmenting
the appropriate models with additional $\cPT$-symmetric fields.
For every conventional scalar universality class, 
there are $\cPT$-symmetric extensions of the model
that exhibit patterning behavior in the vicinity of a critical point. 
Consider a general $\cPT$-symmetric scalar theory in $d$ dimensions.
We suppose there is a set of fields $\phi^a$ that transform trivially under $\mathcal P$ and $\mathcal T$
transformations and a set of fields $\chi^b$ that transform nontrivially in such a way that the action
is invariant under the combined action of the operators $\cPT$.
For example, we can extend the model described by in Eq. (\ref{eq:action}) to
one with an $O\left(N\right)$ symmetry acting on both $\vec\phi$ and $\vec\chi$ fields,
with $\cPT$-symmetric couplings such as $-ig\vec\phi\cdot\vec\chi$.

We find homogeneous equilibrium phases by minimizing $V$,
with $(\phi^a_0,\chi^b_0)$ the global minimum among all homogeneous solutions.
We assume that $\mathcal{PT}$ symmetry is maintained, which implies that $\phi^a_0$ is real,
$\chi^b_0$ is imaginary and $V(\phi_0,\chi_0)$ is real.
The mass matrix $\mathcal{M}$ associated with this solution is given in block form by 
\begin{equation}
\mathcal{M}=\left(\begin{array}{cc}
\frac{\partial^{2}V}{\partial\phi^{2}} & \frac{\partial^{2}V}{\partial\phi\partial\chi}\\
\frac{\partial^{2}V}{\partial\phi\partial\chi} & \frac{\partial^{2}V}{\partial\chi^{2}}
\end{array}\right)
\end{equation}
evaluated at $(\phi^a_0,\chi^b_0)$.
This mass matrix is not necessarily 
Hermitian but is $\mathcal{PT}$-symmetric.
In the two-component case, we had $\mathcal{M} = \sigma_3 \mathcal{M}^* \sigma_3$.
This generalizes to the multicomponent case as
\begin{equation}
\mathcal{M} = \Sigma \mathcal{M}^* \Sigma,
\end{equation}
where $\Sigma$ is a diagonal matrix with entries $\pm 1$.
The characteristic equations for $M$ and $M^*$ are the same, so they have the same eigenvalues. 
As a consequence, the eigenvalues of $\mathcal{M}$ must either both be real or form a complex-conjugate pair.
As before, the zeros of $\det \left( q^2 + \mathcal{M} \right)$ are the poles of the matrix propagator.
In order for the expectation values of the fields $(\phi_0,\chi_0)$ to be stable at quadratic order 
against $q^2=0$ perturbations,
we must have $\det \left(\mathcal{M} \right)>0$. Instability with respect to $q^2=0$ fluctuations
occurs when $\det \left(\mathcal{M} \right)<0$, and $\det \left( q^2 + \mathcal{M} \right)$
has an odd number of real positive roots.
A homogeneous solution represents a normal phase, perhaps metastable, if $\det \left( q^2 + \mathcal{M} \right)$
has zeros only for $q^2$ and negative. The matrix propagator decays exponentially in all channels.
In regions of parameter space where a homogeneous solution gives rise to complex zeros of 
$\det \left( q^2 + \mathcal{M} \right)$, the matrix propagator will exhibit sinusoidal modulation 
of exponential decay in the associated channels, but the homogeneous vacuum is stable.
If a homogeneous solution has $\det \left( \mathcal{M} \right)>0$ and an even number of real, positive roots, 
it will be quadratically stable against $q^2=0$ fluctuations, but not against fluctuations where $q^2$ is a postive
root.
This is the signal that pattern formation is a Lifshitz transition.

The one-loop effective potential $V_{\text{eff}}\left(\phi,\chi\right)$ is given by
\begin{equation}
V_{\text{eff}}\left(\phi,\chi\right)=V\left(\phi,\chi\right)
+\frac{1}{2}  \int \frac{d^d q}{\left( 2\pi\right)^d}
\left[ \log \det \left( q^2 + \mathcal{M} \right) \right].
\end{equation}
When $\det ( q^2 + \mathcal{M})$ is negative for $q^2$ in some positive interval,
the ground state energy density of the homogeneous state $V_{\text{eff}}\left(\phi,\chi\right)$
will have an imaginary part, indicating its instability. This is conventionally interpreted
as a decay rate.
For a state unstable with respect to $q^2=0$ fluctuations, where $\det (\mathcal{M})<0$, the decay
rate is a measure of the time to reach a new phase where the expected values of $\phi$ and $\chi$
are different. For phases exhibiting pattern formation, where $\det (\mathcal{M})>0$, the decay
rate is a measure of the time required to move from a homogeneous phase to the patterned phase,
with no change in the expected values of $\phi$ and $\chi$. In either case, it is given by
\begin{equation}
\Gamma = \frac{\pi}{2}  \int_{\mathcal R} \frac{d^d q}{\left( 2\pi\right)^d},
\end{equation}
where $\mathcal{R}$ is the region of $q$ space where $\det \mathcal{M} < 0$.
This formula is equivalent to the one given by Weinberg and Wu \cite{Weinberg:1987vp},
where the decay rate is given in terms of the zero-point energy of the tachyonic modes.
Note that this decay rate is perturbative, representing a fast decay, in contrast to the slow
modes associated with changing pattern morphology. This indicates that relaxation
from a given initial condition may take little simulation time relative to autocorrelation time.
Adequately sampling the equilibrium state may require a great deal of time depending on the
target observables.
In the general case, there may be more than one homogeneous solution that is unstable to pattern formation.
We cannot necessarily predict which
inhomogeneous phase has the lowest free energy. 
In the case of a theory with three or more components, pattern formation may be quite complicated 
\cite{PhysRevB.84.144402}.

\section{Application to QCD}
%\noindent {\bf Application to QCD.} 
QCD at finite density is a multi-component field theory with a generalized $\mathcal{PT}$ symmetry. Its widely-conjectured phase structure 
is characterized by a first-order line with a critical end point in the $Z(2)$ universality class. Although a thorough treatment of the critical behavior requires an understanding of quark degrees of freedom, it is sufficient for our purposes to focus on the real and imaginary parts $P_R$
and $P_I$ of the trace of the Polyakov loop. The real part, $P_R$, is the order parameter for the pure gauge deconfinement transition. This transition is in the three-dimensional $Z(3)$ universality class, and is first-order.
The order parameter $\left<P_R\right>$ is zero in the low-temperature, confined phase, and jumps at the deconfinement
transition temperature to a nonzero value.
When dynamical quark fields with realistic masses are added at finite temperature, the phase transition is replaced by a rapid
crossover where $\left<P_R\right>$ rapidly increases with temperature. Over the same temperature interval, the chiral condensate of u and d quarks rapidly decreases with temperature.

It is widely believed, mostly on the basis of theoretical models but also simulations, that this crossover behavior 
extends into the $\mu-T$ plane, 
eventually running into a low-temperature, first-order line coming out of the $T=0$ axis, with a second-order critical end point at the point where  the first-order line becomes a cross-over.
It is in the vicinity of the critical end-point that pattern formation is possible.
It is well known that QCD at nonzero temperature and baryon density has a sign problem. The origin
of the sign problem can be seen simply from the contribution of heavy quarks and antiquarks to the
Euclidean space partition function. 
Heavy quarks move through spacetime on essentially straight timelike trajectories.
When a heavy quark traverses a topologically nontrivial closed loop from $t=0$ to $t=\beta=1/T$,
it carries with it a factor of $P$. Similarly, heavy antiquarks carry a factor of $P^+$. This is just
the nonabelian Aharonov-Bohm phase factor for a closed loop.
If $\mu=0$,
these contributions to the partition function are equally weighted, and the partition function is real.
If $\mu\ne 0$, however, nontrivial quark trajectories carry a factor $e^{\beta\mu}P$ and antiquarks
a factor $e^{-\beta\mu}P^\dagger$, and the weights of different configurations in the partition function
become complex.

Euclidean QCD with $\mu\ne 0$ in fact possesses a symmetry of $\cPT$ type; specifically, invariance under $\cC\cK$.
Charge conjugation $\cC$ is an unitary transformation which exchanges $P$ and $P^\dagger$, while
$\mathcal K$ is an anti-unitary transformation which also exchanges $P$ and $P^\dagger$. Although $\mathcal K$
would also conjugate any complex numbers appearing in configuration weights, those do not appear
because they would cause a sign problem at $\mu=0$, which does not occur. Thus QCD in Euclidean space at finite density
has a generalized $\cPT$ symmetry. This is completely consistent with the expected behavior of QCD at finite
density. The expectation value of $P_I$ is zero when the chemical potential is zero. It becomes nonzero
when the chemical potential $\mu$ is nonzero, but $\cC\cK$ symmetry requires it to be purely imaginary. 
This gives rise to two different real expectation values for $P$ and $P^\dagger$. This in turn implies that the free energies required to insert a static quark versus a static antiquark are different, as expected when $\mu\ne 0$.

The universality class of the critical end point is widely believed to be in the $Z(2)$ universality class, which is also
the university class of the conventional liquid-gas phase transition. Because the baryon number density has a discontinuity
across the first-order line, it is a natural expectation.
The three-dimensional effective field theory for $P_R$ and $P_I$ falls naturally into the universality class of the $\cPT$ extension of $Z(2)$ critical behavior, with $P_R$ and $P_I$ identified with $\phi$ and $\chi$ respectively. This raises the possibility that finite-density QCD could exhibit pattern formation near its critical end point, composed of regions of confined and deconfined phase.
This is an attractive and interesting possibility, as it would lead to balls of deconfined quark matter appearing
near the critical end-point on the low-$\mu$ side of the critical line, while balls of confined baryonic matter
would appear on the high-$\mu$ side of the critical line. Rod and sheet phases are also possible.
This is the QCD analog of nuclear pasta in neutron star crusts \cite{Horowitz:2004yf}.
However, this simple picture does not take into account the behavior of the chiral order parameter. We know
from Polyakov-Nambu-Jona Lasinio (PNJL) models and from simulations that analysis of the critical behavior of QCD at finite temperature must take into account
both confinement-deconfinement and chiral phenomena to arrive at a complete description.
It has been suggested that the chiral degrees of freedom may also become unstable
near the critical line due to a Lifshitz instability \cite{Pisarski:2019cvo,Pisarski:2020gkx}. 
Thus we face the exciting possibility that the behavior of QCD at finite density may be
much richer than previously conceived, posing new challenges to theory, simulation and experiment.

\subsection*{Acknowledgements}
MCO thanks C.\ M.\ Bender and Z.\ Nussinov and STS thanks A.\ Grebe and G.\ Kanwar for helpful discussions. S. T. S. was supported by a Graduate Research Fellowship from the U.S. National Science Foundation under Grant No. 1745302; the U.S. Department of Energy, Office of Science, Office of Nuclear Physics, from DE-SC0011090; and fellowship funding from the MIT Department of Physics.

\bibliographystyle{unsrtnat}
\bibliography{stella}

\begin{thebibliography}{38}
\providecommand{\natexlab}[1]{#1}
\providecommand{\url}[1]{\texttt{#1}}
\expandafter\ifx\csname urlstyle\endcsname\relax
  \providecommand{\doi}[1]{doi: #1}\else
  \providecommand{\doi}{doi: \begingroup \urlstyle{rm}\Url}\fi

\bibitem[de~Forcrand(2009)]{deForcrand:2010ys}
Philippe de~Forcrand.
\newblock {Simulating QCD at finite density}.
\newblock \emph{PoS}, LAT2009:\penalty0 010, 2009.
\newblock \doi{10.22323/1.091.0010}.

\bibitem[Gupta(2010)]{Gupta:2011ma}
Sourendu Gupta.
\newblock {QCD at finite density}.
\newblock \emph{PoS}, LATTICE2010:\penalty0 007, 2010.
\newblock \doi{10.22323/1.105.0007}.

\bibitem[Aarts(2016)]{Aarts:2015tyj}
Gert Aarts.
\newblock {Introductory lectures on lattice QCD at nonzero baryon number}.
\newblock \emph{J. Phys. Conf. Ser.}, 706\penalty0 (2):\penalty0 022004, 2016.
\newblock \doi{10.1088/1742-6596/706/2/022004}.

\bibitem[Troyer and Wiese(2005)]{PhysRevLett.94.170201}
Matthias Troyer and Uwe-Jens Wiese.
\newblock Computational complexity and fundamental limitations to fermionic
  quantum monte carlo simulations.
\newblock \emph{Phys. Rev. Lett.}, 94:\penalty0 170201, May 2005.
\newblock \doi{10.1103/PhysRevLett.94.170201}.

\bibitem[Barahona(1982)]{Barahona_1982}
F~Barahona.
\newblock On the computational complexity of ising spin glass models.
\newblock \emph{Journal of Physics A: Mathematical and General}, 15\penalty0
  (10):\penalty0 3241--3253, oct 1982.
\newblock \doi{10.1088/0305-4470/15/10/028}.

\bibitem[Bender and Boettcher(1998)]{Bender:1998ke}
Carl~M. Bender and Stefan Boettcher.
\newblock {Real spectra in non-Hermitian Hamiltonians having PT symmetry}.
\newblock \emph{Phys. Rev. Lett.}, 80:\penalty0 5243--5246, 1998.
\newblock \doi{10.1103/PhysRevLett.80.5243}.

\bibitem[Bender(2007)]{Bender:2007nj}
Carl~M. Bender.
\newblock {Making sense of non-Hermitian Hamiltonians}.
\newblock \emph{Rept. Prog. Phys.}, 70:\penalty0 947, 2007.
\newblock \doi{10.1088/0034-4885/70/6/R03}.

\bibitem[Meisinger and Ogilvie(2013)]{Meisinger:2012va}
Peter~N. Meisinger and Michael~C. Ogilvie.
\newblock {PT Symmetry in Classical and Quantum Statistical Mechanics}.
\newblock \emph{Phil. Trans. Roy. Soc. Lond.}, A371:\penalty0 20120058, 2013.
\newblock \doi{10.1098/rsta.2012.0058}.

\bibitem[Feng et~al.(2017)Feng, El-Ganiany, and Ge]{feng}
Liang Feng, Ramy El-Ganiany, and Li~Ge.
\newblock Non-hermitian photonics based on parity-time symmetry.
\newblock \emph{Nature Photonics}, 11:\penalty0 752--762, 2017.

\bibitem[El-Ganiany et~al.(2018)El-Ganiany, Makris, Khajavikhan, Musslimani,
  Rotter, and Christodoulides]{Christodoulides}
R.~El-Ganiany, K.~G. Makris, M.~Khajavikhan, Z.~H. Musslimani, S.~Rotter, and
  D.~N. Christodoulides.
\newblock Non-hermitian physics and pt symmetry.
\newblock \emph{Nature Physics}, 14:\penalty0 11--19, January 2018.

\bibitem[Miri and Alu(2019)]{Miri}
M.-A. Miri and A.~Alu.
\newblock Exceptional points in optics and photonics.
\newblock \emph{Science}, 363:\penalty0 eaar7709, 2019.

\bibitem[Bender(2019)]{BenderBook}
C.~M. Bender.
\newblock \emph{PT Symmetry: In Quantum and Classical Physics}.
\newblock WorldScientific, 2019.

\bibitem[Fisher(1978)]{Fisher:1978pf}
M.~E. Fisher.
\newblock {Yang-Lee Edge Singularity and phi**3 Field Theory}.
\newblock \emph{Phys. Rev. Lett.}, 40:\penalty0 1610--1613, 1978.
\newblock \doi{10.1103/PhysRevLett.40.1610}.

\bibitem[Ogilvie and Medina(2018)]{Ogilvie:2018fov}
Michael~C. Ogilvie and Leandro Medina.
\newblock {Simulation of Scalar Field Theories with Complex Actions}.
\newblock \emph{PoS}, LATTICE2018:\penalty0 157, 2018.
\newblock \doi{10.22323/1.334.0157}.

\bibitem[Seul and Andelman(1995)]{Seul476}
Michael Seul and David Andelman.
\newblock Domain shapes and patterns: The phenomenology of modulated phases.
\newblock \emph{Science}, 267\penalty0 (5197):\penalty0 476--483, 1995.
\newblock \doi{10.1126/science.267.5197.476}.

\bibitem[Ortix et~al.(2008)Ortix, Lorenzana, and
  Di~Castro]{PhysRevLett.100.246402}
C.~Ortix, J.~Lorenzana, and C.~Di~Castro.
\newblock Coulomb-frustrated phase separation phase diagram in systems with
  short-range negative compressibility.
\newblock \emph{Phys. Rev. Lett.}, 100:\penalty0 246402, Jun 2008.
\newblock \doi{10.1103/PhysRevLett.100.246402}.

\bibitem[Muratov(2002)]{PhysRevE.66.066108}
C.~B. Muratov.
\newblock Theory of domain patterns in systems with long-range interactions of
  coulomb type.
\newblock \emph{Phys. Rev. E}, 66:\penalty0 066108, Dec 2002.
\newblock \doi{10.1103/PhysRevE.66.066108}.

\bibitem[Ravenhall et~al.(1983)Ravenhall, Pethick, and
  Wilson]{Ravenhall:1983uh}
D.~G. Ravenhall, C.~J. Pethick, and J.~R. Wilson.
\newblock {Structure of Matter Below Nuclear Saturation Density}.
\newblock \emph{Phys. Rev. Lett.}, 50:\penalty0 2066--2069, 1983.
\newblock \doi{10.1103/PhysRevLett.50.2066}.

\bibitem[Hashimoto et~al.(1984)Hashimoto, Seki, and Yamada]{10.1143/PTP.71.320}
Masa-aki Hashimoto, Hironori Seki, and Masami Yamada.
\newblock {Shape of Nuclei in the Crust of Neutron Star}.
\newblock \emph{Progress of Theoretical Physics}, 71\penalty0 (2):\penalty0
  320--326, 02 1984.
\newblock \doi{10.1143/PTP.71.320}.

\bibitem[Caplan and Horowitz(2017)]{Caplan:2016uvu}
M.~E. Caplan and C.~J. Horowitz.
\newblock {Colloquium : Astromaterial science and nuclear pasta}.
\newblock \emph{Rev. Mod. Phys.}, 89\penalty0 (4):\penalty0 041002, 2017.
\newblock \doi{10.1103/RevModPhys.89.041002}.

\bibitem[Horowitz et~al.(2004)Horowitz, Perez-Garcia, and
  Piekarewicz]{Horowitz:2004yf}
Charles~J. Horowitz, M.~A. Perez-Garcia, and J.~Piekarewicz.
\newblock {Neutrino - pasta scattering: The Opacity of nonuniform neutron -
  rich matter}.
\newblock \emph{Phys. Rev.}, C69:\penalty0 045804, 2004.
\newblock \doi{10.1103/PhysRevC.69.045804}.

\bibitem[Glaser et~al.(2007)Glaser, Grason, Kamien, Ko{\v{s}}mrlj, Santangelo,
  and Ziherl]{Glaser_2007}
M.~A Glaser, G.~M Grason, R.~D Kamien, A~Ko{\v{s}}mrlj, C.~D Santangelo, and
  P~Ziherl.
\newblock Soft spheres make more mesophases.
\newblock \emph{Europhysics Letters ({EPL})}, 78\penalty0 (4):\penalty0 46004,
  may 2007.
\newblock \doi{10.1209/0295-5075/78/46004}.

\bibitem[Ortix et~al.(2009)Ortix, Lorenzana, and Castro]{ORTIX2009499}
C.~Ortix, J.~Lorenzana, and C.~Di Castro.
\newblock Universality classes for coulomb frustrated phase separation.
\newblock \emph{Physica B: Condensed Matter}, 404\penalty0 (3):\penalty0 499 --
  502, 2009.
\newblock \doi{https://doi.org/10.1016/j.physb.2008.11.045}.

\bibitem[Chaikin and Lubensky(1995)]{cha95}
P.~M. Chaikin and T.~C. Lubensky.
\newblock \emph{Principles of Condensed Matter Physics}.
\newblock Cambridge University Press, Cambridge, 1995.

\bibitem[Stephenson(1970)]{PhysRevB.1.4405}
John Stephenson.
\newblock Ising model with antiferromagnetic next-nearest-neighbor coupling:
  Spin correlations and disorder points.
\newblock \emph{Phys. Rev. B}, 1:\penalty0 4405--4409, Jun 1970.
\newblock \doi{10.1103/PhysRevB.1.4405}.

\bibitem[Giuliani and Seiringer(2016)]{CMP347.983}
A.~Giuliani and R.~Seiringer.
\newblock Periodic striped ground states in ising models with competing
  interactions.
\newblock \emph{Commun. Math. Phys.}, 347:\penalty0 983, 2016.
\newblock URL
  \url{https://link.springer.com/article/10.1007/s00220-016-2665-0}.

\bibitem[Timonin and Chitov(2017)]{PhysRevE.96.062123}
P.~N. Timonin and Gennady~Y. Chitov.
\newblock Infinite cascades of phase transitions in the classical ising chain.
\newblock \emph{Phys. Rev. E}, 96:\penalty0 062123, Dec 2017.
\newblock \doi{10.1103/PhysRevE.96.062123}.

\bibitem[Cross and Hohenberg(1993)]{RevModPhys.65.851}
M.~C. Cross and P.~C. Hohenberg.
\newblock Pattern formation outside of equilibrium.
\newblock \emph{Rev. Mod. Phys.}, 65:\penalty0 851--1112, Jul 1993.
\newblock \doi{10.1103/RevModPhys.65.851}.
\newblock URL \url{https://link.aps.org/doi/10.1103/RevModPhys.65.851}.

\bibitem[Bray(1994)]{doi:10.1080/00018739400101505}
A.J. Bray.
\newblock Theory of phase-ordering kinetics.
\newblock \emph{Advances in Physics}, 43\penalty0 (3):\penalty0 357--459, 1994.
\newblock \doi{10.1080/00018739400101505}.
\newblock URL \url{https://doi.org/10.1080/00018739400101505}.

\bibitem[Langer et~al.(1975)Langer, Bar-on, and Miller]{PhysRevA.11.1417}
J.~S. Langer, M.~Bar-on, and Harold~D. Miller.
\newblock New computational method in the theory of spinodal decomposition.
\newblock \emph{Phys. Rev. A}, 11:\penalty0 1417--1429, Apr 1975.
\newblock \doi{10.1103/PhysRevA.11.1417}.

\bibitem[Binder et~al.(1978)Binder, Billotet, and Mirold]{Binder1978}
K.~Binder, C.~Billotet, and P.~Mirold.
\newblock On the theory of spinodal decomposition in solid and liquid binary
  mixtures.
\newblock \emph{Zeitschrift f{\"u}r Physik B Condensed Matter}, 30\penalty0
  (2):\penalty0 183--195, Jun 1978.
\newblock \doi{10.1007/BF01320985}.

\bibitem[Chakrabarti(1992)]{PhysRevB.45.9620}
Amitabha Chakrabarti.
\newblock Transition from metastability to instability in the dynamics of phase
  separation.
\newblock \emph{Phys. Rev. B}, 45:\penalty0 9620--9625, May 1992.
\newblock \doi{10.1103/PhysRevB.45.9620}.

\bibitem[Chakrabarty and Nussinov(2011)]{PhysRevB.84.144402}
Saurish Chakrabarty and Zohar Nussinov.
\newblock Modulation and correlation lengths in systems with competing
  interactions.
\newblock \emph{Phys. Rev. B}, 84:\penalty0 144402, Oct 2011.
\newblock \doi{10.1103/PhysRevB.84.144402}.

\bibitem[Schmalian and Wolynes(2000)]{PhysRevLett.85.836}
J\"org Schmalian and Peter~G. Wolynes.
\newblock Stripe glasses: Self-generated randomness in a uniformly frustrated
  system.
\newblock \emph{Phys. Rev. Lett.}, 85:\penalty0 836--839, Jul 2000.
\newblock \doi{10.1103/PhysRevLett.85.836}.
\newblock URL \url{https://link.aps.org/doi/10.1103/PhysRevLett.85.836}.

\bibitem[Westfahl et~al.(2001)Westfahl, Schmalian, and
  Wolynes]{PhysRevB.64.174203}
Harry Westfahl, J\"org Schmalian, and Peter~G. Wolynes.
\newblock Self-generated randomness, defect wandering, and viscous flow in
  stripe glasses.
\newblock \emph{Phys. Rev. B}, 64:\penalty0 174203, Oct 2001.
\newblock \doi{10.1103/PhysRevB.64.174203}.
\newblock URL \url{https://link.aps.org/doi/10.1103/PhysRevB.64.174203}.

\bibitem[Weinberg and Wu(1987)]{Weinberg:1987vp}
Erick~J. Weinberg and Ai-qun Wu.
\newblock {Understanding Complex Perturbative Effective Potentials}.
\newblock \emph{Phys. Rev.}, D36:\penalty0 2474, 1987.
\newblock \doi{10.1103/PhysRevD.36.2474}.

\bibitem[Pisarski et~al.(2019)Pisarski, Skokov, and Tsvelik]{Pisarski:2019cvo}
Robert~D. Pisarski, Vladimir~V. Skokov, and Alexei Tsvelik.
\newblock {A Pedagogical Introduction to the Lifshitz Regime}.
\newblock \emph{Universe}, 5\penalty0 (2):\penalty0 48, 2019.
\newblock \doi{10.3390/universe5020048}.

\bibitem[Pisarski et~al.(2020)Pisarski, Rennecke, Tsvelik, and
  Valgushev]{Pisarski:2020gkx}
R.D. Pisarski, F.~Rennecke, A.~Tsvelik, and S.~Valgushev.
\newblock {The Lifshitz Regime and its Experimental Signals}.
\newblock In \emph{{28th International Conference on Ultrarelativistic
  Nucleus-Nucleus Collisions}}, 4 2020.

\end{thebibliography}

\end{document}